\documentclass[5p, times, twocolumn]{elsarticle}
\usepackage{xcolor}
\usepackage{amsmath}
\usepackage{hyperref}

\usepackage{soul}

\newcommand{\makefigure}[2]{
\begin{figure}[ht!]
\begin{center}
\includegraphics[width=0.95\columnwidth]{#1}
\caption{#2 \label{fig:#1}}
\end{center}
\end{figure}}

\newcommand{\makefigurefull}[2]{
\begin{figure*}[!ht]
\begin{center}
\includegraphics[width=.95\linewidth]{#1}
\caption{#2 \label{fig:#1}}
\end{center}
\end{figure*}}

\newcommand{\makefigurestd}[2]{
\begin{figure}[ht!]
\begin{center}
\includegraphics[scale=1]{#1}
\caption{#2 \label{fig:#1}}
\end{center}
\end{figure}}

\newcommand{\makefigurestdcustom}[3]{
\begin{figure}[ht!]
\begin{center}
\includegraphics[scale=#3]{#1}
\caption{#2 \label{fig:#1}}
\end{center}
\end{figure}}

\newcommand{\makefigurefullstd}[2]{
\begin{figure*}[!ht]
\begin{center}
\includegraphics[scale=1]{#1}
\caption{#2 \label{fig:#1}}
\end{center}
\end{figure*}}

\title{Computing Elastic Tensors of Amorphous Materials from First-Principles}

\author[KUL,IMEC]{C.~Pashartis \corref{cor1}}
\author[IMEC]{M.~J.~van~Setten}
\author[KUL]{M.~Houssa}
\author[IMEC]{G.~Pourtois}% \corref{cor2}}
\address[KUL]{Department of Physics and Astronomy, KU Leuven, Leuven, Belgium}
\address[IMEC]{Imec, Leuven, Belgium}
\cortext[cor1]{Corresponding author}
\begin{document}

\begin{frontmatter}

\begin{abstract}

Advancements in modern semiconductor devices increasingly depend on the utilization of amorphous materials and the reduction of material thickness, pushing the boundaries of their physical capabilities. The mechanical properties of these thin layers are critical in determining both the operational efficacy and mechanical integrity of these devices. Unlike bulk crystalline materials, whose calculation techniques are well-established, amorphous materials present a challenge due to the significant variation in atomic topology and their non-affine transformations under external strain. This study introduces a novel method for computing the elastic tensor of amorphous materials, applicable to both bulk and ultra-thin films in the linear elastic regime using Density Functional Theory. We exemplify this method with a-SiO$_2$, a commonly used dielectric. Our approach accounts for the structural disorder inherent in amorphous systems, which, while contributing to remarkable material properties, complicates traditional elastic tensor computation. We propose a solution involving the inability of atomic positions to relax under internal relaxation, near the boundaries of the computational unit cell, ensuring the affine transformations necessary for linear elasticity. This method's efficacy is demonstrated through its alignment with classical Young's modulus measurements, and has potential for broad application in fields such as Technology Computer Aided Design and stress analysis via Raman spectra. The revised technique for assessing the mechanical properties of amorphous materials opens new avenues for exploring their impact on device reliability and functionality.

\end{abstract}


\end{frontmatter}

\bibliographystyle{elsarticle-num}

% insert suggested keywords - APS authors don't need to do this

%\maketitle must follow title, authors, abstract, \pacs, and \keywords
% \maketitle

%-----------------------------------------------------------------------
%
%                       I N T R O D U C T I O N
%
%-----------------------------------------------------------------------

\section{Introduction}

% moved from below

% \textcolor{red}{I think that the introduction needs to be reformulated/reordered and the challenge to model and compute the elastic properties of a disordered system with a periodic boundary software needs to be stressed: atomistic model generation, sampling of enough conformations, derivation of properties accounting from possible  rearrangements of atoms within the model. Why abinitio? versatility of the technique, you don't need to deal with transferability issues of analytical potentials as in classical MD but this is done at the expense of the model size and the use of periodic boundary conditions}

%new

% intro on the use of amorphous materials
Amorphous materials are abundantly integrated into modern semiconductor technologies, assuming a diverse array of roles, from `passive' functions like substrates or electrically insulating layers, to `active' ones, involving charge carrier transport. For example, hydrogenated amorphous silicon (a-Si:H) plays a pivotal role in both photovoltaic solar cells and the control transistors of thin-film liquid crystal displays. This is attributed to its direct band gap, moderate electron mobility, and excellent optical absorption capabilities, see for example Refs. ~\cite{Matsui_PIPRAA_2013, Stuckelberger_RASER_2017}. The anticipated successor, amorphous InGaZnO$_4$ (a-IGZO), employed in high-end display technologies, is reported to offer substantial advantages over traditional silicon-based thin-film transistor materials. Notably, its deposition temperature is compatible with polymeric substrates. These advantages encompass reduced rigidity, improved stability, and enhanced electron mobility, making it feasible to manufacture flexible thin-film transistors for large-area electronic applications. \cite{Nomura_N_2004, Arai__2011}

Furthermore, in Si-based Complementary Metal-Oxide–Semiconductor (CMOS) technology, amorphous silicon dioxide and hafnium dioxide act as dielectric barriers to prevent charge carrier leakage.\cite{Houssa__2003, Wang_CR_2018} Amorphous metals such as Ti based alloys are also frequently employed in gate metals \cite{Matsukawa__2012}, and the effective work function of the device can be adjusted using thin layers of other metals, such as amorphous TiAlC \cite{Kamimuta__2007, Arimura__2021}. Additionally, amorphous bulk materials find extensive use as low-dielectric constant materials to reduce the parasitic capacitance of the CMOS addressing interconnection circuitry or as cost-effective fill materials. \cite{Wu_EJOSSSAT_2014, Grill_APR_2014}

As the transistor's dimensional reduction continues to approach the physical limit, the electronic industry is being driven to transition to three-dimensional integration schemes to sustain performance gains (see Refs.~\cite{Subramanian_2020, Vincent_IJOTEDS_2020}). As a direct consequence the impact of the mechanical properties of material on both the device operability and performances are becoming of paramount interest. The convergence of requirements arising from the device's intended operation and the decreasing thickness of functional material layers necessitates stringent control measures to ensure the structural integrity of these ultra-thin material layers and their associated interfaces. They are also introducing new (or enhanced) sources of stress or strain variability. The effect of strain engineering in layered materials is a well studied topic, with examples including the enhancement of the silicon charge carriers mobility in Si based nanooelectronics \cite{Gupta_NFE_2018,Wang_NE_2018, Jiang_I_2020}. At the atomic level, the dynamics of interfacial effects (strain, defects, etc.) can strongly impact different aspects such as adhesion, charge carrier mobility, band alignment, or the pinning of the Fermi-level.\cite{Poindexter_PISS_1983, Helms_ROPIP_1994, Poindexter_SSAT_1989, Houssa_MB_2009} Our earlier work illustrated that sub-5~nm thick crystalline Ru layers exhibit an increased mechanical flexibility \cite{Pashartis_ASS_2022}. In general, a less stiff elastic tensor has different consequences in terms of device affected properties, which include adhesion of material layers \cite{Sankaran__2018} and enhanced diffusion \cite{Cowern_PRL_1994, Yildiz_MB_2014}. Amorphous materials, due to their intrinsic atomic disordered nature, offer the perspective to accommodate external strain sources differently than their crystalline counterpart would, and are potentially opening new engineering mitigation strategies. For this to happen, there is a need for understanding how the mechanical integrity of amorphous materials is being impacted by thinning. Unfortunately, measuring the elastic properties of a few nanometer thick film is a non-trivial task \cite{Deng_POTNAOS_2009, Amo_AN_2017, Zak_JOMR_2022}, but one that is quantifiable from \textit{ab initio}. An important advantage of computing the elastic tensor through \textit{ab initio} methods is its adaptability for use in a range of simulations and experiments, including applications such as Technology Computer-Aided Design (TCAD) and stress analysis with Raman spectroscopy \cite{Anastassakis_SSC_1993}. In this paper, we report a methodology to compute the elastic tensor of both bulk and ultra-thin film amorphous materials from first-principles. We illustrate the practicality of our approach using amorphous silicon dioxide, a widely utilized dielectric material in Si CMOS technology. 

% More generally, amorphous alloys reflect a non-equilibrium phase from which the characteristic atomic arrangements are close to a liquid one, with no long-range periodicity. They can either be produced by a rapid solidification method that freezes the liquid structure of the alloy melt \cite{Angell_S_1995}, or by other methods (sputtering, bombardment,...) that can mix atoms to achieve a disordered state. 

% section on crystalline vs amorphous
The methodology for numerically evaluating the bulk elastic tensor from \textit{ab-initio} for crystalline materials is widely known, with implementations such as \texttt{ElaStic} \cite{Golesorkhtabar_CPC_2013} and \texttt{Pymatgen} \cite{Jong_SD_2015}. However, there is a large difference when treating an amorphous material model that is not accounted for with their ordered counterpart; the non-affine transformation under external strain and lack of periodicity. In addition, when compared to ordered crystalline systems, amorphous materials are atomically disordered materials; in which, the bond lengths, coordination numbers (CNs), and atomic densities vary wildly throughout the model used. The difficulty is then to build a representation that captures a statically relevant distribution of the atomic disorder, while relying on condensed matter physics formalism with periodic boundary conditions. This typically results in the use of large periodic unit cell models, which can often exceed over 100 atoms, in which atomic disorder is being built. Due to these differences in morphology with respect to ordered systems, different local energetic minima solutions can co-exist. Therefore, modeling amorphous systems requires sufficient sampling of atomic configurations large enough to provide statistically relevant results. In what follows, we give a brief review on the background of amorphous materials, followed by a discussion of the algorithm used to compute the elastic tensor from \textit{ab-initio}, and finally the application to SiO$_2$ bulk and ultra thin films.

\subsection{The Origins of Elasticity and Plasticity of Amorphous Systems}

From a first-principles perspective, the modeling of amorphous materials requires accounting for a large number of atoms to capture any semblance of information that is representative of the macroscopic properties. It is hence computationally difficult to properly assess the mechanical properties. Not only does the atomic configuration change within the same stoichiometric configuration, but the long-range homogeneity and topological heterogeneity can also radically vary. In some instances, the atomic packing density can be very dense, while in others, there could be a formation of sub-lattices within the model (see Ref.~\cite{Wondraczek__2011}). Since many amorphous solids exhibit brittleness, much of the field of mechanical properties of glass has been concerned with determining the fracture or stress yield point in which the material breaks or enters the plastic regime.\cite{Furukawa_NM_2009} These analyses have been performed using a variety of molecular dynamic approaches by simply applying varying normal strain/stress to a material system. The procedure corresponds to deformations normal to the material surface which result in the Young's moduli. However, when considering ultra-thin films or shear deformations of an amorphous system, there is a caveat. 

Due to the disordered nature of bonding in amorphous materials, they are quite heavily prone to changes in atomic position - typically constructing a new topological phase of the initial system. In other words, plastic-like rearrangements of the system are not uncommon. The bond lengths between atoms, the CNs per element, and the topology of voids can all greatly vary depending on the phase and the approach used to build the system. Under external strain, this makes amorphous materials a more ductile alternative to their crystalline counterparts, and one that is capable of reordering and reorganizing under external influence. In fact, metallic glasses have been observed to atomically rearrange under deformation to accommodate strain.\cite{Wang_AM_2012}

% \makefigure{amorphous_topology_towards_ultrastrong_glasses}{Variations of topology prevalent in mid-range structure resulting in varying mechanical properties of glasses from Ref.~\cite{Wondraczek_2011}, \textbf{reused with permission}.}

In crystalline solids, we often deal with elastic and plastic deformations, where the latter is translated by the formation of dislocation defects. The shear stress required to push a system into a plastic regime, where any deformation introduces irreversible changes in shape, is typically referred to as the yield stress. The shear stress is the limit in which the deformation rate ($\dot{\gamma}$) goes to zero \cite{Bonn_ROMP_2017}, 
\begin{equation}
    \sigma_\gamma = \lim_{\dot{\gamma} \to 0} \sigma(\dot{\gamma}) \hspace{0.1cm}.
\end{equation}
In contrast, since there is no recognition of traditional order in amorphous solids, it is difficult to define what a plastic transition is. This is especially so given every amorphous solid of the same stoichiometric composition is unique or a locally stable solution to the configuration of the system. In other words, a plastic deformation realistically yields another amorphous phase of the system. Therefore, any deformation to an amorphous system, whether it be large or small in magnitude, can push the system into a new configuration. Shear stresses on amorphous materials is an ongoing topic of investigation and one which is grounded in Falk and Langer's early work on dynamics of viscoplastic deformation in amorphous solids, where they found that the atomic flow occurs in localized zones or clusters of molecules that shift to release stress \cite{Falk_PRE_1998}. These zones then can cascade local changes in the system, further changing the mechanical properties. In their system, they found linear elasticity up to 2\% of shear strain, after which the onset of the plastic regime occurs. Their modelling work was verified to be observed in slowly sheared systems by Maloney and Lemaitre.\cite{Maloney_PRE_2006} Though both of these systems are two-dimensional, they provided valuable information into the very slow changes that occur at the microscopic level of molecules and atoms, and one which we can observe in our work herein. An extensive review is performed by Bonn \textit{et al.} \cite{Bonn_ROMP_2017} over the current state of theoretical models of shear plasticity and the corresponding yield stress.

With regards to the elastic regime, the glass phase transition temperature ($T_g$) typically has no consistent correlation to the stiffness of the glass, even though its mechanical properties is highly dependent on the shear modulus or viscosity at the transition point. For example, SiO$_2$ and other refractory glasses, have high transition temperatures but are also typically defined by their low atomic packing density.\cite{Rouxel_JOTACS_2007, Wang_PIMS_2012} Metallic glasses on the other hand, have been shown to have their transition temperature correlated with their elastic moduli by $T_g = 2.5 E$ .\cite{Wang_JONS_2005} Even the Young's modulus widely varies from 5~GPa for glassy water to 365~GPa for tungsten-based metallic glasses.\cite{Migliori_PBCM_1993, Ohtsuki_APL_2004} The structural and electronic contribution to the mechanical properties have been suggested to be correlated with the average bond energy and density (or coordination number) \cite{Wondraczek__2011}. Given the variation of interatomic bonding options available to many of the known amorphous solids, the topology or intrinsic structure also contributes. A comprehensive review of various amorphous systems and their elastic properties and values can be found in Ref.~\cite{Wang_PIMS_2012}.

In the linear elastic regime, local changes still occur and compared to crystalline systems, there is a non-affine linear elastic response to deformation. The domination of shear-elastic fluctuations has been heavily reported in theoretical studies \cite{Marruzzo_SR_2013}, suggesting that the bond restructuring and phase changes occur even in the elastic regime. In these models, the particles are typically studied using displacements, forces, model potentials (such as the Leonard-Jones), and frequencies of the sound waves. Though these abstractions don't formally constitute a real system, they yield valuable information, such as the heterogeneity of the local stiffness. The level of disorder (when compared to ordered structures) at the local regime can affect the transverse degrees of freedom of the stiffness resulting in broadening of the energy spectra or the existence of the so-called \textit{boson peak}. \cite{Taraskin_PRL_2001, Chumakov_PRL_2011} This broadening is very clearly observed in Raman Spectra of amorphous solids (see a-Si for example).

The information that even at the atomistic level, the elastic components can vary, in particular, the shear components, leads to the realization that amorphous materials are \textit{extremely good} at reordering bonds to distribute external stresses. In addition, these local elastic variations do not exist at the crystalline level, leading to the conclusion that the elastic tensor of a sufficiently large amorphous material is more akin to a global average of the local variations. Unfortunately, determining the local stiffness is difficult from an \textit{ab-initio} viewpoint as the forces between atoms is not an ingredient for minimization or one that can be gleaned to reconstruct the internal atomic stresses from theorems such as the Virial Theorem. Since we have established the crux of employing a mechanical analysis at the atomic level is due to both the capability of amorphous materials to reorder and construct new phases, these must be taken into account in any elastic tensor generation procedure.

%-----------------------------------------------------------------------
%
%                           M E T H O D S
%
%-----------------------------------------------------------------------
\section{The Amorphous Elastic Tensor Method}

Both experimentally and theoretically (\textit{ab-initio}, molecular dynamics, etc.), one can deform a section of an amorphous solid and measure the stress response of the system, giving the Young's modulus if the stress and strain are parallel. To compute the full elastic tensor, deformations are made to the model unit cell to perturb the system and determine either the change in energy or the change in stress. The former requires the second derivative with respect to the strain applied. In contrast, the latter case is related linearly to the strain applied tends to be more precise,
\begin{equation}
    \vec{\sigma} = \textbf{C} \vec{\epsilon} \hspace{.1cm}.
    \label{eqn:stress_strain_voigt}
\end{equation}
Where $\vec{\sigma}$ is the Voigt representation of the Cauchy stress tensor, $\textbf{C}$ is the fourth-order Elastic tensor, and $\vec{\epsilon}$ is the Voigt representation of the Green-Lagrangian strain tensor. The algorithm from \texttt{pymatgen} \cite{Jong_SD_2015} produces at most \textit{six} unique deformations with \textit{four} varying magnitudes per deformation. The magnitudes for the strains normal to the surface are $[-1\%,~-0.5\%,~0.5\%,~1\%]$ whereas the shear strains follow $[-6\%,~-3\%,~3\%,~6\%]$. These magnitudes are typically sufficient for crystalline systems, but for amorphous systems, one must still ensure to remain in the elastic regime.

As has been reviewed earlier, shear deformations are the linchpin that drive large variations of atomic displacement in amorphous materials. For instance, \autoref{fig:non-held_vs_held_affine_motion}a) illustrates the evolution of the morphology of an amorphous SiO$_2$ model after the minimization of its atomic degrees of freedom with a shear deformation of 3\%. A colour spectrum is provided to demonstrate the amount of normalized displacement occurring per atomic site, accompanied with the direction of displacement as a green vector. It is observed that independent sections of the material shift in differing directions. For example, in the top facet of the unit cell, there is a clockwise-like shift in atomic positions. The resulting elastic tensor for such a system yields varying results since a different applied strain would nudge the system into a different energetic minima. As a result, the elastic stability criteria of Born \cite{Born_MPOTCPS_1940, Mouhat_PRB_2014} is typically not satisfied, leading to negative elastic tensor's eigenvalues and hence to a non-physical result.

\makefigure{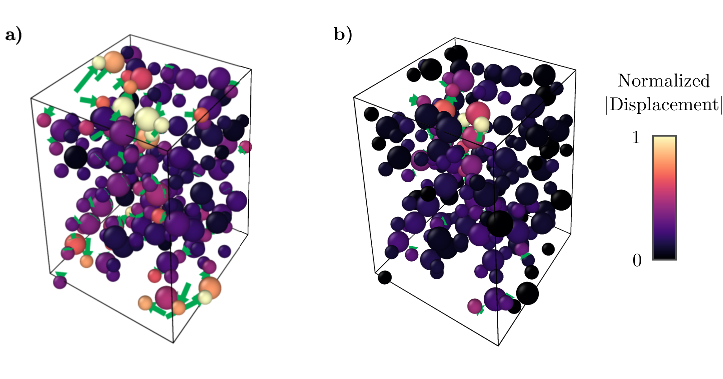}{A unit cell of an SiO$_2$ amorphous alloy under 3\% shear deformation showing atomic displacement after internal relaxation. The first image a) demonstrates the non-affine transformation of atoms to accommodate the deformation. The second image b) is the result of enforcing atomic positions near the unit cell boundary. The colours on the atoms correspond to the normalized magnitude of displacement, with arrows demonstrating the vectors relative to the starting positions. Visualized with \texttt{OVITO} \cite{Stukowski_MASIMSAE_2009}.}

At first glance, one might assume that either i) the applied strain is too large, which perturbs the system out of the linear elastic regime into a plastic one, or ii) that due to amorphous nature of the system, the configuration undoes the applied strain. The consequences on the Voigt-stress components of the elastic tensor in a normally applied direction (\autoref{fig:SiO2_held_non_held_stress_strain}a)). At this stage of the discussion, we are focusing on reproducing the linear elastic response of the material, therefore neglecting any non-linear, viscoelastic, or any other mechanical variant. In this regime, the stress-strain relationship must be strictly \textit{linear}. This assumption is reasonable (but should be used on a per case basis)  as for instance, many glasses, such as silica (a-SiO$_2$), silicates, and aluminates,  have been reported to exhibit elastic properties.\cite{Rouxel_JOTACS_2007} 

The linear elastic regime is clearly broken when observing $\sigma_1$ in \autoref{fig:SiO2_held_non_held_stress_strain}a), the system is responding to the external strain with non-affine transformations yielding a new local structural minimum. Additionally, recall that the relative magnitudes of each curve do not matter, but the variation of the stress is used to evaluate the elastic tensor. In crystalline materials, 1\% strain does not force the material into a plastic regime for the majority of pure ordered materials. Is the non-linearity due to anharmonicity or is it truly due to the plastic limit? Instead of determining either of these options, if case ii) is correct, then the application of elastic tensor methodology to ultra-thin films or surfaces can be applied to amorphous materials.

\makefigurefullstd{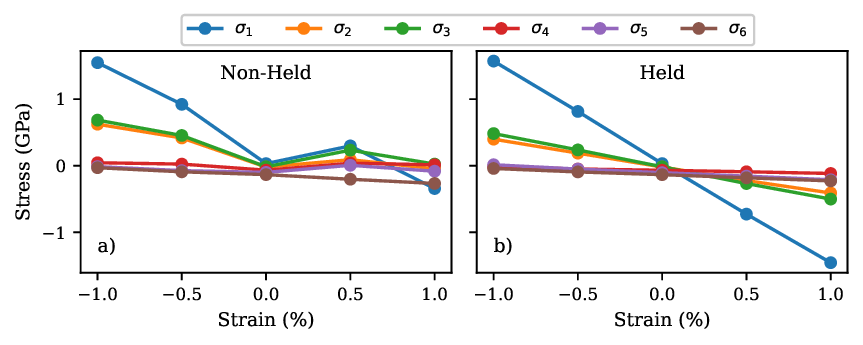}{A comparison of the Voigt-stress components for a normally applied strain for a) straining without constraining atoms and b) constraining the atomic sites within a shell of the unit cell surface. This data is based on the goemetry observed in \autoref{fig:non-held_vs_held_affine_motion}}

If, due to applying strain to the unit cell of an amorphous material, the system undoes the applied strain, then we cannot use the finite perturbation technique in the quasi-harmonic potential limit to determine the mechanics of the material. One of the key-defining features of an amorphous material is its inherent lack of translational symmetry or order, so any enforced deformation to a unit cell without symmetry is therefore capable of reverting to its original state. This is very much the exact same phenomena observed when developing the elastic tensor methodology for ultra-thin films. The solution is to artificially build a region of symmetry that maintains the applied strain. To this end, taking a small volume ($dV$) within the edges of the unit cell periodic boundaries (see \autoref{fig:amorphous_hold_technique}), atoms can be selected to remain invariant under internal relaxation. The internal relaxation of such a scheme can be clearly observed in \autoref{fig:non-held_vs_held_affine_motion}b), where the atoms along the unit cell boundary do not move. The remaining interior atoms still relax their positions, but to a much lesser degree than in \autoref{fig:non-held_vs_held_affine_motion}a). The result is an elastic tensor which typically satisfies the Born criteria for stability. The constraint also yields in a linear response to the applied strain, with respect to stress, as observed comparing the non-held system in \autoref{fig:SiO2_held_non_held_stress_strain}a) to that of the held system in \autoref{fig:SiO2_held_non_held_stress_strain}b). 
%%%%%%%%%%%%%%%%%%%%%%%%
% \newpage
\makefigurestdcustom{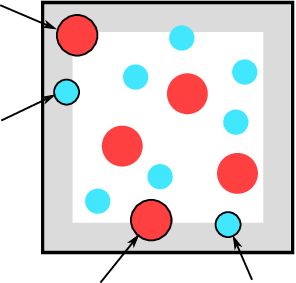}{Schematic diagram of an amorphous solid where a volumetric shell is selected near to the edges of the unit cell periodic boundaries in which to constrain atoms during internal relaxation.}{.75}

The process to calculate an elastic tensor for an amorphous system then simplifies to:
\begin{enumerate}
    \item Relax the constructed amorphous system from a Monte Carlo, molecular dynamics or other scheme
    \item Apply the same deformations for crystalline systems, but enforce a shell of volume near the surface to constrain atomic sites lying within
\end{enumerate}
The distance from the cell edges should be converged such that the number of atoms selected is minimized and the elastic tensor components remain stable and consistent. In practice we found a good metric was to ensure that less than 20\% of the atoms should be constrained and that a very thin shell in the neighbourhood of 0.05-0.25~\r{A} (see Supplementary Material) is sufficient to yield an elastically stable tensor. Of course, general convergence to Density Functional Theory (DFT) parameters should still be carried out. There is no guarantee such a scheme correctly calculates the stiffness of an amorphous material, but by applying an Ackland-Jones analysis \cite{Ackland_PRB_2006} on the local bonding nature of the structure, we observed that by holding certain atoms the symmetry is maintained under strain. Without the technique, the atomic shift ensued could signify that the material is in a non-elastic regime. We instead suppose that that the changes are due to an artifact of the lack of inherent symmetry in the computational unit cell. To assess how valid holding boundary atoms is, a-SiO$_2$ was constructed and the elastic tensor analysis was performed.

To compute the ultra-thin elastic tensor refer to our previous paper ~\cite{Pashartis_ASS_2022}, where the position of the atoms at the surface are held still (which is given for free by the amorphous algorithm). Additionally, the stress tensor must be scaled accordingly using a standardized definition of the thickness of the film which accounts for the electron density extending past its surface.

\section{Computational Details}
\subsection{Generation of Amorphous Structures}

The a-SiO$_2$ models were generated with a \textit{decorate and relax} scheme (see Ref.~\cite{Drabold_TEPJB_2009}). The procedure begins with an atomic structure built from a random distribution of silicon, the atomic position is then compressed and subsequently decorated by oxygen atoms to reproduced coordination numbers similar to the crystalline phase. These models are then optimized using DFT. This methodology has been heavily employed in the study of IGZOs and in general leads to structures with reasonable radial distribution functions and band gaps \cite{Setten_MA_2022}.

A total of ten 96-atom structures and fifteen 200-atom structures were generated to sample the variations in configurations available at densities ranging from 2.3--2.9~$g\cdot cm^{-3}$ (see \autoref{fig:two_variants_SiO2}). Two different sizes of unit cells were studied, a 96-atom cell (Si$_{32}$O$_{64}$) and variants of a 200-atom type cell (ex. Si$_{66}$O$_{132}$). The radial distribution functions of the 200-atom type models for the nearest neighbours and different bond types can be found in \autoref{fig:rdf_sio2}, which reproduce values observed in literature \cite{Temkin_JONS_1975, Ching_PRB_1982, Susman_PRB_1991}.

\makefigure{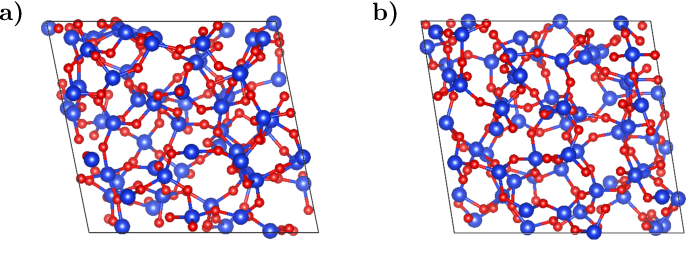}{Two variants of a 198-atom SiO$_2$ with atomic densities of a) 2.6~$g \cdot cm^{-3}$ and b) 2.9~$g \cdot cm^{-3}$, generated with the \textit{decorate and relax} scheme with \texttt{VESTA} \cite{Momma_JOAC_2011}}

\makefigurestd{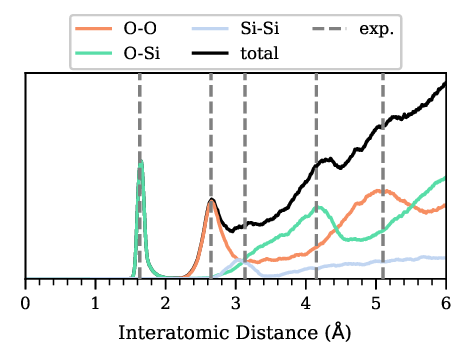}{Radial distribution functions of the 200-atom type models for the O-O, O-Si, and Si-Si bonds, as well as the combined (total) signature. Reference lines are taken from experiment, Ref.~\cite{Susman_PRB_1991}.}

\subsection{Density Functional Theory}
All calculations presented were performed using the first-principle \texttt{CP2K} simulation package, using 3D periodic boundary conditions. A Fermi Dirac distribution for the occupancy of the valence band structure has been imposed with an electronic temperature of 1000~K. The Goedecker-Teter-Hutter \texttt{CP2K} pseudopotential library \cite{Goedecker_PRB_1996, Hartwigsen_PRB_1998, Krack_TCA_2005}, was used and combined with the Perdew–Burke-Ernzerhof (PBE) \cite{Perdew_PRL_1996} exchange correlation functional. DZVP basis sets were used from the \texttt{CP2K} standard library.\cite{VandeVondele_TJOCP_2007} The simulations were converged with respect to the elastic tensor components for the bulk conventional structure to achieve a precision of 10~GPa. The bulk lattice parameters and elastic coefficients were found to be converged with a kinetic energy cutoff of 600~Ry and a relative energy cutoff of 50~Ry (for the Gaussian grid). The maximum force cutoff used for unit cell relaxation was $1\cdot 10^{-4}$~bohr$^{-1}$~Ha. A single k-point was used to perform the calculation since amorphous materials have no Bravais lattice due to lack of symmetry. Convergence of DFT parameters was performed on a single amorphous system. The unit cells were minimized to reach their optimum lattice vector orientation and position.

\subsection{Additional Details}
The distance from the unit cell edges for the boundary atoms was converged to select the minimum number of atoms, at a distance of 0.3~\r{A} from the edge. For example, in the SiO$_2$ system with 198 atoms, on average, 11\% of all the atoms in the system were constrained after the deformations were applied.

The ultra-thin SiO$_2$ films were constructed with the relaxed amorphous bulk variants, oriented with the direction of thickness aligning with the Cartesian-z axis. They were given a vacuum height of 20~\r{A}. The films were allowed to relax while maintaining the same lattice vector directions but not magnitudes, using same tolerances as the bulk before performing the necessary deformations. The atoms at the edge were constrained, combined with the amorphous methodology of constraining atoms near the boundaries of the unit cell.

For the definition of the thickness of the film, at the time of the research, the more precise thickness fitting to the Hartree potential (see Ref.~\cite{Pashartis_ASS_2022}) was not implemented, therefore these results are reported with the vacuum removed. Note that the difference is observed to be minimal. As a consequence, the atomic densities are then reported without thickness correction and are strictly the edge-to-edge or vacuum removed for the z-axis of the volume.

Note that for the orientation of the elastic tensor, both in the bulk and ultra-thin film structures, the first lattice vector is colinear with the Cartesian x-axis. While in the bulk, the topological homogeneity implies that the lattice vector orientations will not significantly matter, assuming a well-distributed amorphous configuration and a large enough configuration space was sampled for stiffness. The selection of the third lattice vector in the direction of the Cartesian-z for ultra-thin films allows those structures and their elastic tensor components in that direction to be easily compared without worry of a change of coordinates.

Coordination numbers were analyzed using the \texttt{pymatgen} \texttt{CrystalNN} method, which utilizes Voronoi analysis to ascertain the neighbourhood of atoms to select for bond pairs.\cite{Ong_CMS_2013} A recent study by Pan \textit{et al.} 2021 \cite{Pan_IC_2021} suggests it to be one of the best frameworks consistent with experimental evidence of coordination numbers.

\section{Results \& Discussion}
\subsection{Bulk a-SiO$_2$}
The mechanical properties of a-SiO$_2$ are widely varied due to the plethora of methods in which to synthesize it, such as the deposition process, and varying temperatures or pressures in the growth in environment. Its characterization produces differing results, which is understood to be due to the differences in structural configuration natural to amorphous materials. Depending on the strength of the covalent bonds, the orientations, their number, the coordination of each atom, and the number and locations of voids, the mechanical properties can be expected to change. The variation in the density and its associated shear, bulk, and Young Hills' moduli are shown in \autoref{fig:SiO2_comparison_Bondi}. For reference, densities close to 2.2 $g\cdot cm^{-3}$ are used in devices. The best-fit was computed for the data and tends towards the values previously computed with \textit{ab-initio} from Ref.~\cite{Bondi_PRB_2010}, without the elastic tensor. Experimental values of a-SiO$_2$ have found the Young's modulus to be 70~GPa \cite{Rouxel_JOTACS_2007} for bulk silica and 76.6$\pm$7.2~GPa \cite{Ni_APL_2006} for nano wires, measured at room temperature. Extrapolated to 0~K, the experimental Young's modulus for bulk silica is near 65~GPa (note the density from this reference is not reported), placing the trends of our results and those of other DFT calculations within good accuracy of experimental data.

A linear best-fit line was computed to demonstrate the general trend of the data, as it is too sparse to fully determine if the relation is linear or non-linear. The R$^2$ values of each line of fit ranges between 0.55-0.56, but a $\chi_{red}^2$ cannot be computed due to the lack of a quantity of variance. Traditionally, the Young's modulus and the atomic density are given on a log-log plot, typically called a material selection chart (see Ref.~\cite{Shercliff__2016}), but this does not imply that any material is non-linear with respect to the density and modulus. For amorphous materials, there have been reports of linear scaling with the two properties, one example being Harms \textit{et al.} 2003 \cite{Harms_JONS_2003}, where different annealing temperatures generated amorphous structures of different densities and stiffness. Therefore, any best-of-fit lines reported in this amorphous work should be considered general trends and not necessarily validation of linear behaviour.

\makefigurestd{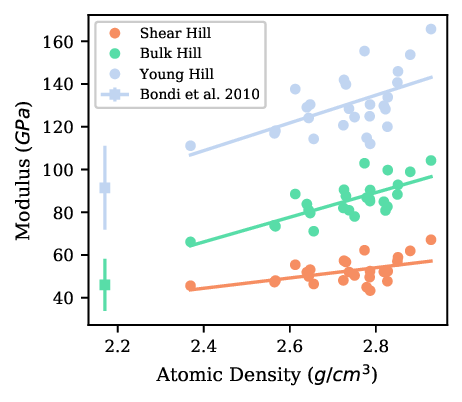}{Comparison of the computed amorphous structures' shear, bulk, and Young's Hill's moduli with \cite{Bondi_PRB_2010} of bulk a-SiO$_2$. The vertical line on the reference data suggests the variation expected at that density. The best-fit of the computed structures is given ($R^2$ values are between 0.55-0.56 and are meant as a guide).}

The values reported by Bondi \textit{et al.} 2010 for a-SiO$_2$ consist of up to 100-atom amorphous configurations where the moduli were computed using the standard technique of straining in a particular direction (or isotropically) and the subsequent stress response is measured. Namely, the Young's modulus is typically computed as
\begin{equation}
    Y = \frac{\sigma_{xx}}{\epsilon} \hspace{0.1cm},
\end{equation}
with $Y$ being the Young's modulus, $\sigma_{xx}$ the stress in the \textit{xx} component (or \textit{11}), and $\epsilon$ the magnitude of strain in the \textit{x} direction.
In contrast, the Young's modulus for this work uses the components of the elastic tensor,
\begin{equation}
    Y = \frac{K_{VRH} \cdot G_{VRH}}{3 K_{VRH} + G_{VRH}} \hspace{0.1cm},
\end{equation}
where $K_{VRH}$ and $G_{VRH}$ are the Voigt-Reuss-Hill average bulk modulus and shear modulus. The shear components of the elastic tensor are included in this formulation, suggesting that the methodology applied to compute the tensor is within the variation of existing literature. At the higher densities, we observe a variation of approximately $\pm 20$~GPa for the Young's modulus and $\pm 10$ GPa for the bulk modulus. This range of stiffness is also observed by Bondi \textit{et al.} \cite{Bondi_PRB_2010} (with a lower density). They suggested, when compared to a-Si, that the addition of H or O to the system weakens the bonds, creating a more deformable system, while reducing the coordination numbers, demonstrating another key-defining attribute of less stiff amorphous systems.
%, performed with the Perdew-Wang exchange correlation functional using \texttt{VASP}.

\subsubsection{Analysis of Bonds and Coordination}
In an effort to determine the source of the changing elastic properties for a given density, a coordination analysis was performed on each structure's relaxed state. The variation of the average coordination number with the Young's modulus in \autoref{fig:young_hill_CN} suggests that as the density is increased, the coordination number typically increases. This is observed from the colours becoming lighter as the density increases. Except for one or two coordination pairs, the majority of the pairs are Si-O. The result coincides with the literature that suggests that given any atom, the more bonds it has (higher CN), then the more resistant it should be to external stresses, thereby stiffening the material system. \cite{Wondraczek__2011} Much like a simple spring system or ball and chain model of atoms, the more connections added to an atom, the more difficult it is to perturb it. The result does assume that the chemical environment, or strength of the bonds, remains approximately the same throughout each configuration computed. The change in average CN is rather small, on the order of 0.15 at the ends of the spectra. Within a small unit of density, the average CN does indeed change non-uniformly in some cases, but the overall trend towards higher stiffness and increased average CN remains.

To determine whether bond direction enforces stiffness in SiO$_2$, the CN result was further post-processed to yield the pseudo-bonds or vectors to the nearest neighbours, each vector was unique to avoid double counting of the same pair (only one of Si-O or O-Si). After taking the element-wise absolute values of the vector, it can then be projected onto the Cartesian axis to determine the portion of the bond in the $\hat{x}, \hat{y},\hat{z}$ directions. Enforcing that each component is positive, is necessary to ensure that each vector adds constructively and not destructively which would yield cancellation. If these vectors are normalized to ignore bond length bias, then averaged, the resulting gradient in \autoref{fig:avg_bond_distribution_sio2_project_norm} and \ref{fig:std_bond_distribution_sio2_project_norm} is observed (colour scale). In particular, the averaged and standard deviation normalized bond direction components are shown with the $c_{11}, c_{22}, c_{33}$ (\textit{xx, yy, zz}) elastic tensor components over the atomic densities. We note that in a sufficiently statistically sampled and large enough amorphous material such as SiO$_2$, the mechanical properties are expected to be isotropic. Since the $c_{11}$ and $c_{22}$ components have similar best-fit trends, but the $c_{33}$ component is different, this highlights the difficulty in sampling enough topologies and varied densities to properly determine the mechanical properties of amorphous materials.

\makefigurestd{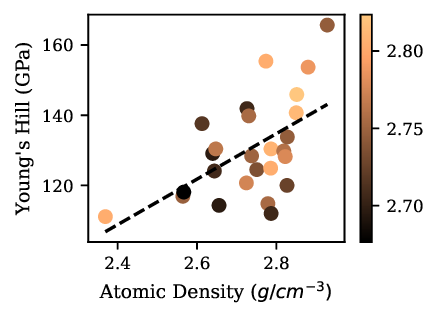}{Young's Hill modulus of bulk a-SiO$_2$ as a function of the atomic density and the average coordination number (coloured spectrum) for each amorphous structure. The dotted line represents the best-fit of the data.}

In \autoref{fig:avg_bond_distribution_sio2_project_norm}, there does not appear to be correlation over the spectrum of densities computed with regards to the elastic tensor component and the pseudo-bond in the same direction. If the average bond strength remains similar across each structure over every atom, one would expect the respective component of the tensor to be stiffer with more bonds along the same direction. To observe this, for any small width of atomic density, the data below the best linear fit should be darker in colour than those above. This suggests that either insufficient structures have been used to properly gather the necessary statistics, or that the direction of the bonds has no effect on the macroscopic stiffness of the amorphous material. The former being the obvious choice since amorphous materials restructure and compensate for external sources of stress.

\makefigurefullstd{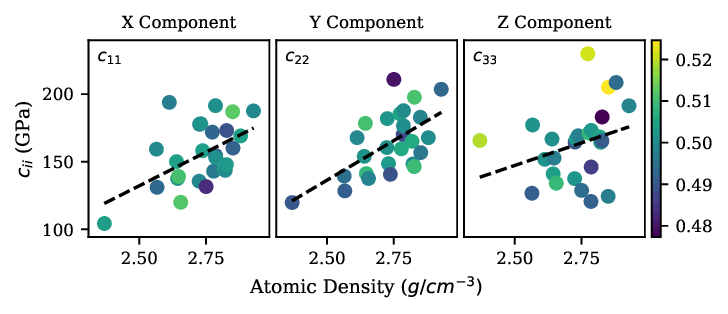}{Projection of averaged X, Y, Z normalized bond direction components (coloured spectrum) compared with their respective elastic tensor components for bulk a-SiO$_2$.}

The variation in average bond direction along each component is very small among the calculated structures, inferring that the amorphous configurations is well disordered. The standard deviation of the bond direction within each structure is shown in \autoref{fig:std_bond_distribution_sio2_project_norm}. The values suggest standard deviations on the order of 2\% of the projection of the pseudo-bond components, which is substantially small, further confirming that the amorphous structures are sufficiently disordered. The lack of variation does indeed suggest that the average pseudo-bond directions for each structure do not affect the final macroscopic stiffness of the material. To further support this observation, the average bond length variation over the densities computed is $10^{-3}$~\r{A}, which is within numerical precision. Meaning that variations for the Si-O bonds within a given structure are extremely small or negligible.

\makefigurefullstd{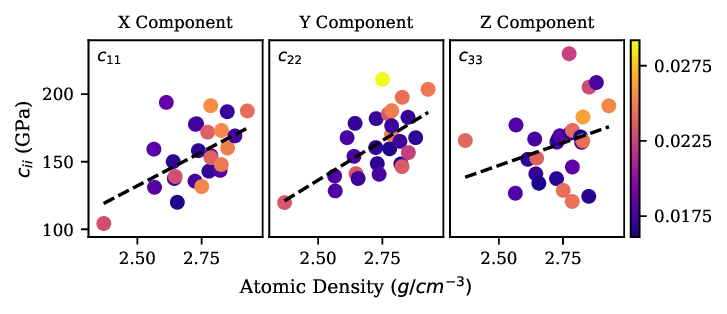}{Projection of standard deviation in the X, Y, Z normalized bond direction components (coloured spectrum) compared with their respective elastic tensor component for bulk a-SiO$_2$. Directions were enforced such that they add constructively.}

\subsection{Ultra-thin a-SiO$_2$}
\subsection{Results \& Discussion of Ultra-Thin Amorphous Silicon Dioxide}
What has been discerned thus far is the dependency of SiO$_2$ on both the bond strength and the coordination number. Extending such a study to ultra-thin silicon oxide is valuable due to its prevalence in CMOS transistors, where they are typically on the order of sub-2~nm. Decoupling the dielectric layer to study individually as a free-standing film enables us to study its individual mechanical properties, which is presented here. At the time of this research, it is not clear if mechanical properties of ultra-thin SiO$_2$ have been computed or derived experimentally.

Combining the approach of ultra-thin film and the amorphous elastic tensor methodologies, the elastic components of the different amorphous configurations (from their corresponding bulk structure) are calculated in \autoref{fig:SiO2_amorphous_thickness_thesis}. The films after cell optimization range from 1-1.6~nm in thickness (surface to surface distance). The statistical averages are shown as the large dots, with the values of the many configurations given by the smaller, lighter dots. A guiding line is drawn between these averages to demonstrate the expected decrease for any configuration, on average, of a given tensor component. It is evident that, like the metallic and crystalline ultra-thin film systems, the most affected component in magnitude is that of $c_{33}$ of 50\%- the one related to stress or strain applied normally to the surface. However, both the $c_{13}$ (43\%) and the $c_{23}$ (48\%) components demonstrate large percentage deviations relative to the bulk average, as observed in \autoref{tab:comparison_film_bulk_comp}. Which are responsible for the in-plane stiffness due to a normally applied stress. Any deviations of the tensor components with smaller magnitudes should be considered carefully, as the precision of the calculation is on the order of 10~GPa, representing a 20\% potential error. Note that due to selecting the thickness to be the surface-to-surface distance, the values of the elastic tensor will migrate to a larger thickness (rightward on the figure) and decrease in stiffness by the equivalent amount of thickness added. This is due to the stress being inversely proportional to any elastic tensor component.

% \makesidewaysfigure{SiO2_amorphous_thickness_thesis}{Elastic tensor components of ultra-thin a-SiO$_2$. Large dots represent the statistical average of the configurations (small dots), with the range of variations given by the crosshairs. A best-fit line is drawn with the average of the configurations to demonstrate the trend.}{4_2:sio2_thin_elastic}
\makefigurefull{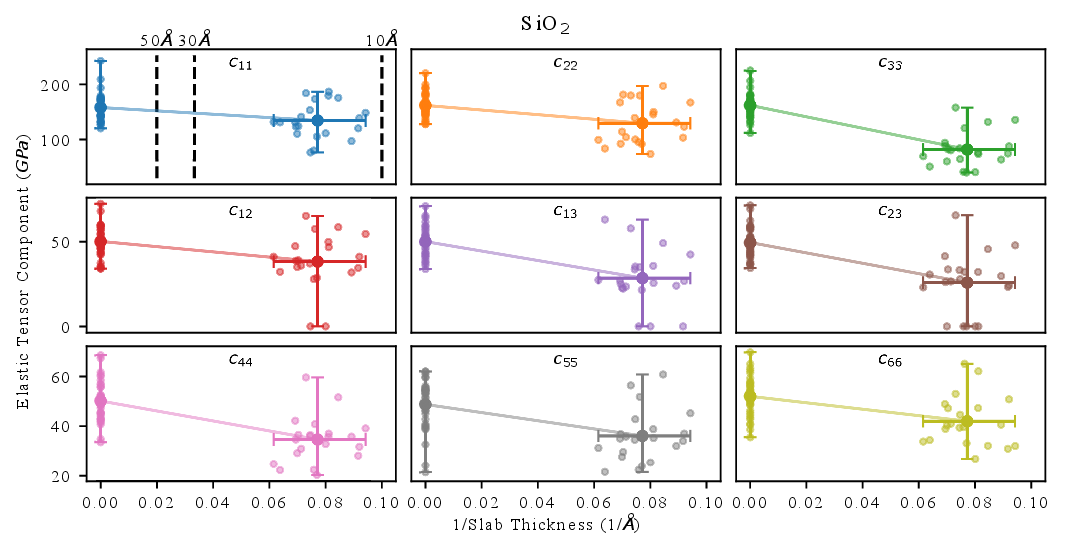}{Elastic tensor components of ultra-thin a-SiO$_2$. Large dots represent the statistical average of the configurations (small dots), with the range of variations given by the crosshairs. A best-fit line is drawn with the average of the configurations to demonstrate the trend.}

\begin{table}[!ht]
% table 1
\centering
% \resizebox{\textwidth}{!}{%
\begin{tabular}{lcccc}
\hline \hline
         & Avg. Bulk (GPa)      & Avg. Film (GPa)      & \% $\Delta$ \\
$c_{11}$ & 158 & 134 & -15  \\
$c_{22}$ & 161 & 129 & -20  \\
$c_{33}$ & 162 & 81  & -50  \\
$c_{12}$ & 50  & 38  & -24  \\
$c_{13}$ & 50  & 28 & -43  \\
$c_{23}$ & 49 & 26 & -48    \\
$c_{44}$ & 50  & 34  & -31  \\
$c_{55}$ & 48  & 36  & -26  \\
$c_{66}$ & 52  & 42  & -19  \\
\hline \hline
\end{tabular}%
% }
\caption{Average values of the configurations of the bulk and ultra-thin films of a-SiO$_2$, ranging from 1-1.6~nm, elastic tensor components. The percent variation from the bulk ($\Delta$) is reported.}
\label{tab:comparison_film_bulk_comp}
\end{table}

In ultra-thin film systems (see Ref.~\cite{Pashartis_ASS_2022}), dangling bonds at the surface contribute to the change in stiffness; increased strain at the surface was observed relative to where the surface would be if it were fully coordinated. Such an analysis cannot be performed easily for amorphous thin films, since the atomic configuration is widely varied, and the atoms drastically reorder themselves to accommodate for dangling bonds. Studying the change in average coordination of the top-most and bottom-most atoms within 1~\r{A} from the average coordination of the entire film suggests that indeed this is also the case in amorphous ultra-thin films between 1-1.6~nm. The change in coordination number was found to be on average -0.9 with an associated standard deviation of 0.3, meaning that compared to the entirety of the ultra-thin film, there are on average atoms with \textit{one} less bond at the surface. Thus it can be considered that the surface has dangling bonds. Additionally, the films were found to have expanded on average 8\% times (standard deviation of 6\%) in thickness when compared to its initial starting position from the fully coordinated equivalent. Combined, the lowering of the coordination at the surface and the stretching of the system in the direction of vacuum, suggest that even in amorphous ultra-thin films (compared to crystalline ones \cite{Pashartis_ASS_2022}), the change in stiffness is due to the dangling bonds and to the relaxed degrees of freedom along the film confined direction. The introduction of the vacuum breaks the homogeneity found in-plane throughout the bulk amorphous state, creating a more deformable system out-of-plane. We expect that at larger thicknesses, the surface effects will be overwritten by the contribution from the centre of the film body, outweighing the effect of the dangling bonds and surface relaxation, thereby increasing the stiffness of the system towards the bulk values.

\section{Conclusion}
%-----------------------------------------------------------------------
%
%                       C O N C L U S I O N
%
%-----------------------------------------------------------------------

As more amorphous materials are being included in semiconductor device stacks, particularly so as miniaturization continues, the need for multiscale properties increases to properly simulate and predict these systems. This paper has addressed the need to calculate the elastic tensor of amorphous solids, required to understand the effect of strain in various subjects such as thermally induced strain,  strain-enhanced diffusion, or adhesion of layers. We have reviewed the origins and discussed the plastic and elastic regimes of amorphous materials. Given that a material can be described in the linear elastic limit, a methodology to compute the elastic tensor has been shown to be in agreement with literature for a-SiO$_2$. The methodology can be extended to any amorphous system in the linear elastic regime, but the efficacy should still be evaluated. Due to the nature of amorphous materials, small changes in external strain often result in a change the topological phase, often leading to plastic transformations. Since the \textit{ab-initio} methodology requires internal relaxation after externally applied strain, this often leads to a sudden change in topology, breaking the affine transformation requirement of the linear elastic regime. To accommodate this issue, a thin shell of atoms around the boundaries of a computational unit cell are selected to remain stationary during internal atomic relaxation. It was found that for a small statistical sample of bulk a-SiO$_2$, the average coordinate number combined with the atomic density is correlated with the Young's Hill modulus. Additionally, there is no nearest neighbour directional component to the corresponding direction of elastic tensor component (ex. $c_{11}$ with the Cartesian X direction). For ultra-thin film a-SiO$_2$ constructed from the bulk counterpart, it was confirmed that similarly to crystalline materials, the dangling bond or electron density at the surface contributes to a reduced stiffness in the direction perpendicular to the surface.

\section{Data availability}
Pre-density functional theory relaxed unit cells of the bulk a-SiO$_2$ have been deposited in the Cambridge Crystallographic Data Centre (CCDC) online database under the deposition numbers of 2313547-2313596, which can also be found searching for this publication.

\section{Acknowledgements}
The resources and services used in this work were partly provided by the VSC (Flemish Supercomputer Center), funded by the Research Foundation - Flanders (FWO) and the Flemish Government. The various members of our group for discussions and aid over the course of this research.

%-----------------------------------------------------------------------
%
%                       B I B L I O G R A P H Y
%
%-----------------------------------------------------------------------

\bibliography{amorphous.bib}

\clearpage

\end{document}